\documentclass[twocolumn,nofootinbib,preprintnumbers,superscriptaddress,amsmath,amssymb,PRL]{revtex4}
\usepackage{graphicx}
\usepackage{gensymb}
\usepackage{dcolumn}
\usepackage{float}
\usepackage{mathptmx, courier, pifont}
\usepackage[scaled=0.92]{helvet}
\usepackage[T1]{fontenc}
\usepackage{textcomp}
\usepackage{color}
\usepackage[normalem]{ulem}

\usepackage[dvipsnames]{xcolor}
\usepackage[colorlinks=true,urlcolor=Blue,linkcolor=Blue]{hyperref}
\usepackage[all]{hypcap}
\usepackage{makecell}
\usepackage{booktabs}
\usepackage{multirow}

\begin{document}

%\linenumbers 

%\title{High-precision measurement of  $Q$-value for neutrino mass determination}

\title{Direct high-precision measurement of the mass difference of $^{77}$As--$^{77}$Se\\ related to neutrino mass determination}
%confirming $^{77}$As a possible $\beta$-decay candidate to determine the neutrino mass
%Comment: is "rules out" too stronly said? In any case the title could be shorter. 
\author{Z.~Ge}\thanks{Corresponding author: zhuang.z.ge@jyu.fi}
\affiliation{Department of Physics, University of Jyv\"askyl\"a, P.O. Box 35, FI-40014, Jyv\"askyl\"a, Finland}%
\author{T.~Eronen}\thanks{Corresponding author: tommi.eronen@jyu.fi}
\affiliation{Department of Physics, University of Jyv\"askyl\"a, P.O. Box 35, FI-40014, Jyv\"askyl\"a, Finland}%
\author{M.~Ramalho}\thanks{Corresponding author: madeoliv@jyu.fi}
\affiliation{Department of Physics, University of Jyv\"askyl\"a, P.O. Box 35, FI-40014, Jyv\"askyl\"a, Finland}%
\author{A.~de Roubin}
%\affiliation{Department of Physics, University of Jyv\"askyl\"a, P.O. Box 35, FI-40014, Jyv\"askyl\"a, Finland}%
\affiliation{Centre d'Etudes Nucl\'eaires de Bordeaux Gradignan, UMR 5797 CNRS/IN2P3 - Universit\'e de Bordeaux, 19 Chemin du Solarium, CS 10120, F-33175 Gradignan Cedex, France}%
\affiliation{KU Leuven, Instituut voor Kern- en Stralingsfysica, B-3001 Leuven, Belgium}%
\author{D.~A.~Nesterenko}
%\email[]{dmitrii.nesterenko@jyu.fi}
\affiliation{Department of Physics, University of Jyv\"askyl\"a, P.O. Box 35, FI-40014, Jyv\"askyl\"a, Finland}%
\author{A.~Kankainen}
\affiliation{Department of Physics, University of Jyv\"askyl\"a, P.O. Box 35, FI-40014, Jyv\"askyl\"a, Finland}%
\author{O.~Beliuskina}
\affiliation{Department of Physics, University of Jyv\"askyl\"a, P.O. Box 35, FI-40014, Jyv\"askyl\"a, Finland}%
%\author{C.~Delafosse}
%\affiliation{Department of Physics, University of Jyv\"askyl\"a, P.O. Box 35, FI-40014, Jyv\"askyl\"a, Finland}%
\author{R.~de~Groote}\thanks{Present address: KU Leuven, Instituut voor Kern- en Stralingsfysica, B-3001 Leuven, Belgium}
\affiliation{Department of Physics, University of Jyv\"askyl\"a, P.O. Box 35, FI-40014, Jyv\"askyl\"a, Finland}%
%Sarina Geldhof
\author{S.~Geldhof}\thanks{Present address: GANIL, CEA/DSM-CNRS/IN2P3, Bd Henri Becquerel, 14000 Caen, France}%
%\thanks{Present address: KU Leuven, Instituut voor Kern- en Stralingsfysica, B-3001 Leuven, Belgium}
%\footnote{Present address: KU Leuven, Instituut voor Kern- en Stralingsfysica, B-3001 Leuven, Belgium}
%\thanks{Present address: KU Leuven, Instituut voor Kern- en Stralingsfysica, B-3001 Leuven, Belgium}
\affiliation{Department of Physics, University of Jyv\"askyl\"a, P.O. Box 35, FI-40014, Jyv\"askyl\"a, Finland}%
%\altaffiliation[]{Present address: KU Leuven, Instituut voor Kern- en Stralingsfysica, B-3001 Leuven, Belgium}
%\alsoaffiliation{lab 2}  \altaffiliation{}
%\altaffiliation{Current address: Institut2 woanders}
%\affiliation[Institut1]{Institut1, Stadt, Land}
\author{W.~Gins}
\affiliation{Department of Physics, University of Jyv\"askyl\"a, P.O. Box 35, FI-40014, Jyv\"askyl\"a, Finland}%
\author{M.~Hukkanen}
\affiliation{Department of Physics, University of Jyv\"askyl\"a, P.O. Box 35, FI-40014, Jyv\"askyl\"a, Finland}%
%\affiliation{University of Bordeaux,....?}
%\affiliation{Centre d\’Etudes Nucl\´eaires de Bordeaux Gradignan, UMR 5797 CNRS/IN2P3 - Universit\´e de Bordeaux, 19 Chemin du Solarium, CS 10120, F-33175 Gradignan Cedex, France}
\affiliation{Centre d'Etudes Nucl\'eaires de Bordeaux Gradignan, UMR 5797 CNRS/IN2P3 - Universit\'e de Bordeaux, 19 Chemin du Solarium, CS 10120, F-33175 Gradignan Cedex, France}
%Jukka Jaatinen
%\author{J.~Jaatinen} 
%\affiliation{Department of Physics, University of Jyv\"askyl\"a, P.O. Box 35, FI-40014, Jyv\"askyl\"a, Finland}%
\author{A.~Jokinen} 
\affiliation{Department of Physics, University of Jyv\"askyl\"a, P.O. Box 35, FI-40014, Jyv\"askyl\"a, Finland}%
\author{\'A.~Koszor\'us}\thanks{Present address: KU Leuven, Instituut voor Kern- en Stralingsfysica, B-3001 Leuven, Belgium}%
%{Á. Koszorús}
\affiliation{Department of Physics, University of Liverpool, Liverpool, L69 7ZE,  United Kingdom}%
%Kotila, Jenni-Mari
\author{J.~Kotila}
\affiliation{Finnish Institute for Educational Research, University of Jyv\"askyl\"a, P.O. Box 35, FI-40014, Jyv\"askyl\"a, Finland}%
\affiliation{Center for Theoretical Physics, Sloane Physics Laboratory Yale University, New Haven, Connecticut 06520-8120, USA}
\affiliation{International Centre for Advanced Training and Research in Physics, P.O. Box MG12, 077125 Bucharest-M\u{a}gurele, Romania}%
%>1 Finnish Institute for Educational Research, University of Jyväskylä, P.O. Box 35, 40014 >Jyväskylä, Finland
%>2 Center for Theoretical Physics, Sloane Physics Laboratory Yale University, New Haven, >Connecticut 06520-8120, USA
\author{J.~Kostensalo}
\affiliation{Natural Resources, Natural Resources Institute Finland, Yliopistokatu 6B, FI-80100 Joensuu, Finland}%
%\affiliation{Department of Physics, University of Jyv\"askyl\"a, P.O. Box 35, FI-40014, Jyv\"askyl\"a, Finland}%
%\author{L.~Canete}
%\affiliation{Department of Physics, University of Jyv\"askyl\"a, P.O. Box 35, FI-40014, Jyv\"askyl\"a, Finland}%
%\altaffiliation{Present address: Aalto University, P.O. Box 11000, FI-00076 Aalto, Finland}
%\author{H.~Penttil\"a}
%\affiliation{Department of Physics, University of Jyv\"askyl\"a, P.O. Box 35, FI-40014, Jyv\"askyl\"a, Finland}%
%\author{I.~Pohjalainen}
%\affiliation{Department of Physics, University of Jyv\"askyl\"a, P.O. Box 35, FI-40014, Jyv\"askyl\"a, Finland}%
\author{I.~D.~Moore}
\affiliation{Department of Physics, University of Jyv\"askyl\"a, P.O. Box 35, FI-40014, Jyv\"askyl\"a, Finland}%
\author{P.~Pirinen}
\affiliation{Department of Physics, University of Jyv\"askyl\"a, P.O. Box 35, FI-40014, Jyv\"askyl\"a, Finland}%
\author{A.~Raggio}
\affiliation{Department of Physics, University of Jyv\"askyl\"a, P.O. Box 35, FI-40014, Jyv\"askyl\"a, Finland}%
\author{S.~Rinta-Antila}
\affiliation{Department of Physics, University of Jyv\"askyl\"a, P.O. Box 35, FI-40014, Jyv\"askyl\"a, Finland}%
%\author{V.~A.~Rubchenya}
%\affiliation{Department of Physics, University of Jyv\"askyl\"a, P.O. Box 35, FI-40014, Jyv\"askyl\"a, Finland}%
%\author{M.~Vilen}
%\affiliation{Department of Physics, University of Jyv\"askyl\"a, P.O. Box 35, FI-40014, Jyv\"askyl\"a, Finland}%
%\author{M.~Vil\'en}
%\affiliation{Department of Physics, University of Jyv\"askyl\"a, P.O. Box 35, FI-40014, Jyv\"askyl\"a, Finland}%
%Vasile Alin Sevestrean ovidiu.nitescu@nipne.ro
\author{V.~A.~Sevestrean}
\affiliation{International Centre for Advanced Training and Research in Physics, P.O. Box MG12, 077125 Bucharest-M\u{a}gurele, Romania}%
\affiliation{Faculty of Physics, University of Bucharest, 405 Atomiștilor, P.O. Box MG11, 077125 Bucharest-M\u{a}gurele, Romania}%
\affiliation{“Horia Hulubei” National Institute of Physics and Nuclear Engineering, 30 Reactorului, POB MG-6, RO-077125 Bucharest-M\u{a}gurele, Romania}
\author{J.~Suhonen}\thanks{Corresponding author:  jouni.t.suhonen@jyu.fi}%
\affiliation{Department of Physics, University of Jyv\"askyl\"a, P.O. Box 35, FI-40014, Jyv\"askyl\"a, Finland}%
\affiliation{International Centre for Advanced Training and Research in Physics, P.O. Box MG12, 077125 Bucharest-M\u{a}gurele, Romania}%
\author{V.~Virtanen}
%Virtanen, Ville
\affiliation{Department of Physics, University of Jyv\"askyl\"a, P.O. Box 35, FI-40014, Jyv\"askyl\"a, Finland}%
%\author{I.~D.~Moore} 
%Andrew  Weaver
%Sasha 	Zadvornaya
%Andrea Raggio
%\author{A.~P.~Weaver}
%\affiliation{School of Computing, Engineering and Mathematics, University of Brighton, Brighton BN2 4JG, United Kingdom}
\author{A.~Zadvornaya}\thanks{Present address: University of Edinburgh, Edinburgh, EH9 3FD, United Kingdom}
\affiliation{Department of Physics, University of Jyv\"askyl\"a, P.O. Box 35, FI-40014, Jyv\"askyl\"a, Finland}%
%I.D. Moore
%\author{J.~\"Ayst\"o}
%\affiliation{Department of Physics, University of Jyv\"askyl\"a, P.O. Box 35, FI-40014, Jyv\"askyl\"a, Finland}%
%%%%%%%%%%%%%University of Jyv\"askyl\"a, Department of Physics, P.O. Box 35, FI-40014 University of Jyv\"askyl\"a, Finland

%A. de Roubin, T. Eronen, O. Beliushkina, L. Canete,
%R. de Groote, A. Jokinen, M. Hukkanen, A. Kankainen, J. Kostensalo,
%I.D. Moore, D. Nesterenko, S. Rinta-Antila, J. Suhonen,
%A. de Roubin ,* J. Kostensalo , T. Eronen , L. Canete ,† R. P. de Groote , A. Jokinen , A. Kankainen ,
%D. A. Nesterenko , I. D. Moore , S. Rinta-Antila , J. Suhonen , and M. Vil´en ‡ Sami Rinta-Antila

%\pacs{21.10.Dr, 27.40.+z, 29.20.db}% 26.30.Ca,
%\pacs{21.10.Dr, 27.40.+z, 29.20.db}% 26.30.Ca,  Atomic Mass  Evaluation 2020 (AME2020) $Q$-value
%ground-state to ground-state electron-capture
\date{\today}
\begin{abstract}
The first direct determination of the ground-state-to-ground-state ${\beta^{-}}$-decay $Q$-value of $^{77}$As to $^{77}$Se was performed by measuring their atomic mass difference utilizing the double Penning trap mass spectrometer, JYFLTRAP. The resulting $Q$-value is 684.463(70) keV, representing a remarkable 24-fold improvement in precision compared to the value reported in the most recent Atomic Mass Evaluation (AME2020). With the significant reduction of the uncertainty of the ground-state-to-ground-state $Q$-value and knowledge of the excitation energies in $^{77}$Se from $\gamma$-ray spectroscopy, the ground-state-to-excited-state $Q$-value of the transition $^{77}$As (3/2$^{-}$, ground state)
$\rightarrow$ $^{77}$Se$^{*}$ (5/2$^{+}$, 680.1035(17) keV) was refined to be 4.360(70) keV. We confirm that this potential low $Q$-value ${\beta^{-}}$-decay transition for neutrino mass determination is energetically allowed at a confidence level of about 60$\sigma$. 
Nuclear shell-model calculations with two well-established effective Hamiltonians were used to estimate the partial half-life for the low $Q$-value transition. The half-life was found to be of the order of 10$^{9}$ years for this first-forbidden non-unique transition, which rules out this candidate a potential source for rare-event experiments searching for the electron antineutrino mass.
%this candidate feasible for neutrino mass searches in a long term measurement. 
%beta-decay half-life of 187Re to be (4.23 ± 0.13) · 1010 y
%alongside an accurate phase-space factor and a statistical analysis of the log(ft) values of known allowed ${\beta^{-}}$-decays, 
%~\cite{Suhonen1998,Avignone2008,Ejiri2019}
\end{abstract}
%\pacs{21.10.Dr, 21.30.-x, 27.60.+j}
\maketitle
\section{Introduction}
The discovery of neutrino oscillations has challenged the idea of massless neutrinos, thus requiring extension of the Standard Model to account for neutrino mass~\cite{Fukuda1998,SNOCollaboration2002,Gerbino2018a}. Neutrino oscillations provide the possibility to infer the neutrino-mass splittings to a high precision. However, they are not sensitive to the absolute neutrino-mass scale, e.g., the mass of the lightest mass eigenstate. To access the absolute mass scale, three complementary approaches are currently being explored: neutrinoless double-${\beta^{}}$ decay, cosmological observations, and kinematic studies of weak-interaction processes such as single ${\beta^{\pm}}$ decays and electron capture (EC)~\cite{Suhonen1998,Avignone2008,Ejiri2019,Velte2019,FORMAGGIO2021}. Among them, high-precision measurements of single ${\beta^{\pm}}$ decays or EC are considered to be the most model-independent methods to determine the absolute scale of the (anti)neutrino mass,
%~\cite{Drexlin2013,Aker2019,Gastaldo2014,Gastaldo2017,Alpert2015,Faverzani2016,Croce2016}. 
as they require no prior assumption on the basic nature (Dirac vs. Majorana) of the neutrino.

%-----------------------------Fig. 1 --------------------------------
% simultaneously, , where the ions were produced and transported with an He gas flow and electric fields. 
\begin{figure*}[!htb]
\centering
\includegraphics[width=1.99\columnwidth]{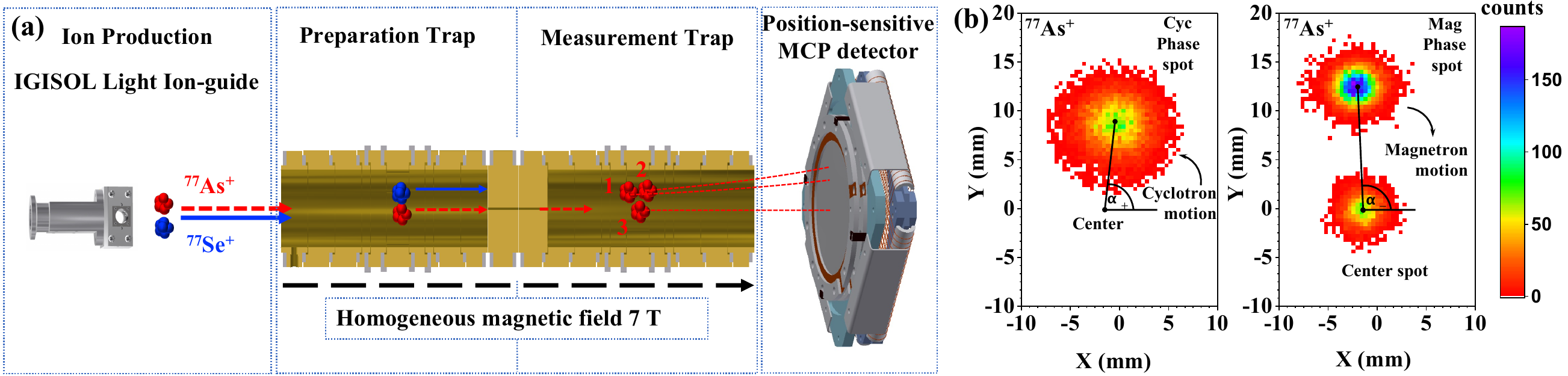}
\caption{(Color online).  (a) Schematic view of ion production and mass measurements using the PI-ICR technique at IGISOL. The $^{77}$As$^+$ and $^{77}$Se$^+$ ions were produced from the fusion reaction with a 9-MeV deuteron beam from K130 cyclotron, bombarding a germanium target of approximately 2~mg/cm$^2$ in thickness.
%and the offline glow-discharge ion source.To produce the ions of interest at IGISOL, a thin germanium target with a thickness of about 2~mg/cm$^2$ was bombarded with a 9-MeV deuteron beam from the K-130 cyclotron. The reaction simultaneously produced both  $^{77}$As$^+$ and  $^{77}$Se$^+$ ions.
Ions having mass number of 77 were selected with a dipole magnet  and transported to the JYFLTRAP PTMS for final ion species selection in the  preparation trap by means of a buffer-gas cooling technique and cyclotron frequency determination using the phase-imaging technique in the measurement trap. A position-sensitive MCP detector was used to register the images of the motion  phases. 
(b) An illustration of the radial-motion (``magnetron'', ``cyclotron'', and ``center'') projection of the $^{77}$As$^+$ ions onto the position-sensitive MCP detector. 
The cyclotron phase spot is displayed on the left side and the magnetron phase spot on the right. The angle difference between the two spots relative to the center spot is utilized to deduce the cyclotron frequency of the measured ion. The color bar indicates the number of ions in each pixel.
%Ion spots (center, cyclotron phase and magnetron phase) of $^{75}$As$^{+}$ on the 2-dimensional position-sensitive MCP detector after a typical PI-ICR excitation pattern with an accumulation time of 400 ms. 
} 
\label{fig:igisol-scheme}
\end{figure*}
%-----------------------------Fig. 1 --------------------------------
%
%

The neutrino mass can be determined by comparing the measured and the expected endpoint energy ($Q_{\beta}$). The presence of a finite neutrino mass causes a shift in the decay endpoint and alters the shape of the decay spectrum, particularly in the proximity of the endpoint. 
%A finite neutrino mass will shift the decay endpoint and modify the decay spectrum shape in the endpoint region.  Relative to the rate, it is most prominent in the close vicinity of the endpoint. 
For $\beta$ decays, the fraction of decay events that fall into an energy interval just below the endpoint energy is proportional to $Q_{\beta }^{-3}$~\cite{Ferri2015}, while for EC the proportionality of the event fraction to $Q_{\beta}$ can be even steeper.
This implies that nuclear transitions with the lowest possible $Q$-value are desirable~\cite{Ferri2015}. Currently,  two nuclei $^{3}$H ($\beta^-$ decay) and $^{163}$Ho~\cite{Eliseev2015,Ranitzsch2017} (EC), with low ground-state-to-ground-state {(gs-to-gs)}  decay $Q$ values of 18.59201(7) keV ~\cite{Myers2015}, and 2.8632(6) keV~\cite{Eliseev2015,Ranitzsch2017,schweiger2024}, are employed for long-term direct neutrino-mass measurements~\cite{FORMAGGIO2021} in experiments such as KATRIN (KArlsruhe TRitium Neutrino)~\cite{Aker2019,Aker2022}, ECHo (Electron Capture in $^{163}$Ho)~\cite{Gastaldo2014,Gastaldo2017,Velte2019,Echo2023}, Project 8~\cite{Project8-23} and HOLMES~\cite{Nucciotti2018,HOLMES23}. 

%correction: 
%Currently, three nuclei $^{3}$H, $^{187}$Re~\cite{Nesterenko2014,Aker2022} ($\beta^-$ decay) and $^{163}$Ho~\cite{Eliseev2015,Ranitzsch2017} (EC), with low ground-state-to-ground-state (gs-to-gs) decay $Q$-values of 18.59201(7) keV, 2.4709(13) keV~\cite{Filianin21}, and 2.833(30)$_{\rm stat}$(15)$_{\rm sys}$ ~\cite{Eliseev2015,Ranitzsch2017}, are employed for long-term direct neutrino-mass measurements~\cite{Nucciotti2012,Ferri2015} in experiments such as KATRIN (KArlsruhe TRitium Neutrino), MARE (Microcalorimeter Arrays for a Rhenium Experiment), and ECHo (Electron Capture in $^{163}$Ho)~\cite{Gastaldo2014,Gastaldo2017,Velte2019}. 
%%%%%changed to 
%Currently,  two nuclei $^{3}$H ($\beta^-$ decay) and $^{163}$Ho~\cite{Eliseev2015,Ranitzsch2017} (EC), with low ground-state-to-ground-state (gs-to-gs) decay $Q$-values of 18.59201(7) keV, and 2.833(30)$_{\rm stat}$(15)$_{\rm sys}$ ~\cite{Eliseev2015,Ranitzsch2017}, are employed for long-term direct neutrino-mass measurements~\cite{Nucciotti2012,Ferri2015} in experiments such as KATRIN (KArlsruhe TRitium Neutrino), ECHo (Electron Capture in $^{163}$Ho)~\cite{Gastaldo2014,Gastaldo2017,Velte2019}, Project 8~\cite{Project8-23} and HOLMES~\cite{HOLMES23}. 
%

%2.492(30)$_{\rm stat}$(15)$_{\rm syst}$  keV~\cite{Basunia2017,Nesterenko2014}, Basunia2017,
% crucial  to desirable 
Further explorations for nuclear $\beta$-decay or EC transitions with low $Q$-values are desirable for prospective (anti)neutrino mass determination experiments~\cite{Mustonen2010,Mustonen2011,Haaranen2013,Suhonen2014}. One transition of interest is the $\beta^-$ transition $^{77}$As (3/2$^+$, 38.790(50) h) $\rightarrow$ $^{77}$Se$^*$ (5/2$^-$, $E^*=$ 680.1035(17) keV) due to its small gs-to-es state $Q^*$ ($Q$ - $E^*$) value of 3.1(17) keV. The $Q^*$ value can be deduced from the high-precision excitation-energy $E^*$ evaluation ~\cite{NNDC}, with sub-keV precision, and the gs-to-gs $Q$ value of 683.2(17) keV from AME2020~\cite{Wang2021,Huang2021}, which provides the main uncertainty of $Q^*$. The gs-to-gs $Q$ value of $^{77}$As in AME2020 is evaluated from  reaction and $\beta^{-}$-decay data of $^{80}$Se(p,$\alpha$)$^{77}$As, $^{76}$Ge($^{3}$He,d)$^{77}$As, and $^{77}$As($\beta^{-}$)$^{77}$Se, with influence of 32\%, 31.8\%, and 17.9\%, respectively.

%Direct $Q$-value measurements of the candidate $\beta$-decay transition in $^{77}$As are strongly desired. Accurate determination of the  $Q$ value, is crucial to  predict the shape of the spectrum and assess the fraction of decay events that falls into the energy interval just below the end-point energy for the electron neutrino mass determination.  %Potential low $Q$-value $\beta$-decay candidates have recently been intensively studied via Penning-trap mass spectrometry 
Recent and thorough investigations into possible low $Q$-value $\beta^-$-decay and EC candidate transitions have employed the Penning-trap mass spectrometry (PTMS) technique~\cite{Sandler2019,Karthein2019a,DeRoubin2020,ge2021,ge2021b,ERONEN2022,Ge2022a,Ge2022b,Ramalho2022,Gamage22,Keblbeck2023,Redshaw2023,Ge2023}, highlighting the importance of high-precision direct $Q$-value measurements prior to utilisation of these nuclei in long-term (anti)neutrino-mass measurements. These earlier investigations have revealed significant discrepancies (exceeding 10 keV) between $Q$ values obtained through indirect methods, such as reaction and decay spectroscopy, compared to those derived from direct mass measurements. This inconsistency spans a wide range of mass numbers, see e.g. Refs.~\cite{Nesterenko2019,ge2021,Ge2023}. 

In this work, we report on the first direct gs-to-gs ${\beta^{-}}$-decay $Q$-value measurements of $^{77}$As with the JYFLTRAP double PTMS at the University of Jyv\"askyl\"a~\cite{Eronen2012,Kolhinen2013,Moore2013}. The $Q$-value of the transition to the excited state of interest is evaluated and determined with a high precision. Moreover, the half-life of the candidate transition is assessed based on nuclear shell-model calculations in order to validate the possibility of using $^{77}$As for future long-term antineutrino-mass determination experiments.
%------------------------------------------------------------
\section{Experimental description}
The direct measurement of the $Q$ value, based on mass difference measurements of the decay pair ions of $^{77}$As$^+$ and  $^{77}$Se$^+$, was carried out at the Ion Guide Isotope Separator On-Line facility (IGISOL) using the JYFLTRAP double Penning trap mass spectrometer~\cite{Eronen2012}, situated at the University of Jyv\"askyl\"a~\cite{Moore2013,Kolhinen2013}. 
To produce the ions of interest at IGISOL, a 9-MeV deuteron beam from the K-130 cyclotron was directed onto a thin germanium target with a thickness of approximately 2~mg/cm$^2$.  Ions of $^{77}$As$^+$ and  $^{77}$Se$^+$ were simultaneously produced through fusion-evaporation reaction.
These produced ions were stopped in the gas cell of the IGISOL light-ion ion guide~\cite{Huikari2004} by colliding with high-purity helium gas at a pressure of about 100 mbar. During this process, the highly charged ions undergo recombination, predominantly adopting a singly charged state. The resulting recoils exited the gas cell through a small nozzle into a sextupole ion guide (SPIG)~\cite{Karvonen2008}, which transports the ions into high vacuum and a subsequent electrode system accelerates them to an energy of 30 keV.
%The beam of ions was then steered with an electrostatic kicker and guided through a.
A dipole magnet with a mass resolving power of approximately 500 adequately isolated only the isobaric ions with $A = 77$, where $A$ represents the mass number. Following the separation, the mass-number-selected ions were transported through an electrostatic switchyard housing a fast kicker electrode used to chop the beam for an optimum number of ions. 
%Subsequently, after passing through the switchyard, the ions were
The ions were then injected into a radiofrequency quadrupole (RFQ) cooler-buncher~\cite{Nieminen2001}, which cooled and bunched the beam.  Finally, the resulting bunches were transported to the JYFLTRAP double Penning trap for the actual frequency ratio measurement.
%A magnetic dipole mass separator with a mass resolving power of about 500 is sufficient to reject all but $A = 77$ ions, where $A$ is the mass number. After the separation, the mass number selected ions are transported through an electrostatic switchyard housing a fast kicker electrode used to chop the beam to have an optimum number of ions. After the switchyard the ions are injected into a radiofrequency quadrupole (RFQ) cooler-buncher~\cite{Nieminen2001}, which is used to cool and bunch the beam. 
%, consisting of two cylindrical traps in a superconducting 7 T magnet.   

The JYFLTRAP double Penning trap comprises two cylindrical traps located in a  7-T superconducting magnet, as illustrated schematically in Fig.~\ref{fig:igisol-scheme}(a). The cooled and bunched ions are confined using the combination of a homogeneous magnetic field and a quadrupolar electrostatic potential inside the traps, where they undergo the superposition of three simple harmonic modes, one axial and two radial. The first trap, designated as the preparation trap, serves as a high-resolution mass separator. In contrast, the second trap, referred to as the measurement trap, is utilized for finely detailed mass determination using ion-cyclotron-resonance techniques.
%The first trap, the purification trap, functions as a high-resolution mass separator, while the second trap, the precision trap, is employed for high-precision mass determination through ion-cyclotron-resonance techniques.
%using either the conventional time-of-flight ion-cyclotron-resonance (TOF-ICR) method~\cite{Koenig1995,Graeff1980} or the phase-imaging ion-cyclotron-resonance (PI-ICR) technique~\cite{nesterenko2021,Nesterenko2018,Eliseev2014,Eliseev2013}.

The ion beam contained $^{77}$Ge$^+$ as a co-produced impurity.
In the first trap, an isobarically purified sample of ions was prepared using the mass-selective buffer gas cooling method~\cite{Savard1991}, providing a typical resolving power $M/\Delta M \approx 10^{5}$.
%The ions of interest will be centered in the trap   by applying first a dipole excitation and then a quadrupole excitation. After  collisions with the buffer gas, only the ions near the center of the trap will be extracted through a pumping barrier of 1.5 mm in diameter separating the first and the second  trap.
%To obtain a clean sample of  $^{77}$Se$^+$ or $^{77}$As$^+$, the mass-selective buffer gas cooling method was sufficient to remove isobaric contaminants and any other ion species present in the beam.
This method was sufficient to provide a clean sample of $^{77}$Se$^+$ or $^{77}$As$^+$ ions.

The determination of the gs-to-gs $Q$ ($Q_{\beta^-}^0$) value is based on the measurement of the cyclotron frequency,
% \begin{equation}
%\label{eq:nuc}
%$\nu_{c}=\frac{1}{2\pi}\frac{q}{m}B$, 
$\nu_{c}=\frac{1}{2\pi}\frac{q}{m}B$, 
%\end{equation} 
%where B is the magnetic field strength, $Q$ is the charge and m the mass of the ion, is determined in the second Penning trap.
where   $q/m$ is the charge-to-mass ratio of the stored ion and $B$ is the magnetic field strength, for both the decay parent and decay daughter ions. 
In this work, the phase-imaging ion-cyclotron-resonance (PI-ICR) technique~\cite{nesterenko2021,Nesterenko2018} was employed for measuring the cyclotron frequencies.
%, offering a speed about 25 times faster to reach a certain precision compared to the conventional time-of-flight ion-cyclotron-resonance (TOF-ICR) method~\cite{Koenig1995,Graeff1980} method. 
%This technique depends on projecting the ion motion in the Penning trap onto a position-sensitive multichannel-plate (MCP) ion detector. micro-channel plate (MCP)
%These patterns are otherwise identical except for the switching on instant of the 
Specifically, measurement scheme number 2 described in~\cite{Eliseev2014} was utilized for directly measuring the cyclotron frequency.

Two timing patterns, denoted as ``magnetron'' and ``cyclotron'',  were employed (see Refs.~\cite{Nesterenko2018,nesterenko2021} for detailed information). These patterns were nearly identical, differing only in the initiation moment of the $\pi$-pulse that converts the ions' cyclotron motion to magnetron. In the ``magnetron'' pattern, ions primarily circulated in the trap with magnetron motion for a duration $t_{acc}$ (accumulation time), while in the ``cyclotron'' pattern, ions revolved with cyclotron motion. These patterns produced magnetron and cyclotron spots or phases on the position-sensitive micro-channel plate (MCP) detector located after the trap, as illustrated in Fig.~\ref{fig:igisol-scheme}(a). 
%In the ``magnetron'' pattern the ions predominantly revolve in the trap for a time duration $t_acc$ (accumulation time) with magnetron motion while in the ``cyclotron'' pattern the ions revolve with cyclotron motion. The exact knowledge of the switch-on time difference $t_acc$ is essential. The used patterns produce so-called magnetron and cyclotron spots or phases on the position-sensitive micro-channel plate (MCP) detector.%~\cite{PS-MCP}.
%are utilized to measure the magnetron or cyclotron motion phases after  free revolution for a same duration t for a phase accumulation.
%
The patterns enable the calculation of the angle between the phases of cyclotron and magnetron motion with respect to the center spot, 
%\begin{equation}
%\label{eq:alphac}
   $\alpha_c = \alpha_+-\alpha_-$, 
%\end{equation}
where $\alpha_+$ and $\alpha_-$ are the polar angles of cyclotron and magnetron phases, respectively. The acquired ``magnetron'' and ``cyclotron'' phase positions of $^{77}$As$^+$ ions are depicted in the left and right panels of Fig.~\ref{fig:igisol-scheme}(b).  It is essential to determine the position of the motional center spot, also shown in Fig.~\ref{fig:igisol-scheme}(b). 

Finally, the cyclotron frequency $\nu_{c}$  is deduced from: 
\begin{equation}
\label{eq:nuc2}
%\nu_{c}=\frac{\alpha_{c}+2\pi(n_{+} + n_{-})}{2\pi{t}},
%\nu_{c}=\frac{\alpha_{c}+2\pi (n_{+} + n_{-})}{2\pi{t}}.
\nu_{c}=\frac{\alpha_{c}+2\pi n_{c}}{2\pi{t}}, 
% m=r(m_{ref}-m_e)+m_e,
\end{equation}
where $n_{c}$ is the number of complete revolutions of the ions during the phase accumulation time $t_{acc}$.
The measurement procedure ensured that $\alpha_c$ remained small to reduce systematic shifts caused by image distortion. To achieve this, the duration $t_{acc}$ was selected to closely align with integer multiples of the $\nu_c$ period. %Accurate knowledge of this switch-on time difference  $t_{acc}$ was crucial. 
This minimized the angle $\alpha_c$, ensuring minimal impact from the interconversion of magnetron and cyclotron motions~\cite{Eliseev2014,Kretzschmar2012c}. During the measurements, $\alpha_c$ did not exceed a few degrees.
%The measurement is set up so that $\alpha_c$ will be small in order to minimize systematic shifts due to image distortion by choosing $t_acc$ to be as close to integer-multiples of $\nu_c$ period as possible.
%$n_{+}$ and  $n_{-}$

Two different accumulation times (330 ms and  321 ms) for both  $^{77}$Se$^+$ and  $^{77}$As$^+$ ions during the measurement were used. The chosen duration $t_{acc}$ ensured the separation of the cyclotron spot of the target ion from any potential isobaric, isomeric, or molecular contaminants. The delay of the initiation of the $\pi$ pulse was repeatedly scanned over one magnetron period and the final extraction delay was varied over one cyclotron period to account for any residual magnetron and cyclotron motion that could shift the different spots. This constituted in a total of $5\times5=25$ scan points for both magnetron and cyclotron phase spots. 
%As the decay parent   $^{77}$Se$^+$ and the decay daughter  $^{77}$As$^+$ ions were produced simultaneously at IGISOL, a direct doublet measurement was realized after separation and purification of the samples. 
%As the decay parent   $^{77}$Se$^+$ and the decay daughter  $^{77}$As$^+$ ions were produced simultaneously at IGISOL and were alternatively measured in JYFLTRAP after separation and purification of the samples, a direct doublet measurement was realized.

%Simultaneously generated at IGISOL, the decay parent $^{77}$Se$^+$ and decay daughter $^{77}$As$^+$ ions underwent alternating measurements in JYFLTRAP following sample separation and purification. This process facilitated a direct doublet measurement. 
%The measurement of the $\nu_{c}$ of the ions  $^{77}$Se$^+$ and  $^{77}$As$^+$  was performed continuously for a total duration of about 13 hours.%
% The following is coming up in the next chapter
%switching between the two every 4 complete scans lasting about three minutes.

The $Q^0_{\beta^-}$ value is directly obtained through the cyclotron frequency ratio, 
%\begin{equation}
    $R = {\nu_{c,Se}}/{\nu_{c,As}}$, 
       % $R = \frac{\nu_{c,Se}}{\nu_{c,As}}$, 
%\end{equation}
where $\nu_{c,As}$ is the cyclotron frequency for $^{77}$As$^+$ and $\nu_{c,Se}$ for  $^{77}$Se$^+$.
During this experiment, alternating measurements of the $^{77}$As$^+$ and  $^{77}$Se$^+$ cyclotron frequency measurements were conducted every few minutes to minimize the contribution of magnetic field fluctuations in the measured cyclotron frequency ratio. Still, 
%for both the decay parent and decay daughter ions strength $B$ was calibrated  by measuring the cyclotron frequency of a reference ion with a well-known mass value before and after measurement of the ion of interest.
linear interpolation was employed to determine the magnetic field at the moment of the parent cyclotron frequency measurement.
%The frequency measurement directly yields the $Q^0_{\beta^-}$-value (see Eq.~\ref{eq:Qec}) via the cyclotron frequency ratio $R$
The $Q^0_{\beta^-}$ value, mass difference of $^{77}$As and $^{77}$Se, can be given as: 
 \begin{equation}
\label{eq:Qec}
%Q^0_{\beta^-} = (R-1)(M_d - m_e)c^2+\Delta{B_{m,d}},
Q_{\beta^-}=(M_p - M_d)c^2 = (R-1)(M_d - qm_e)c^2+(R \cdot B_{d} - B_{p}),
% m=r(m_{ref}-m_e)+m_e,Q_{\beta^-}
\end{equation}
where $M_p$ and  $M_d$ are the atomic masses of the parent  ($^{77}$As) and daughter  ($^{77}$Se), respectively.  $m_{e}$ denotes the mass of an electron, and $R$ represents the cyclotron frequency ratio ($\frac{\nu_{c,d}}{\nu_{c,p}}$) for singly charged ions ($q=1$).  The electron binding energies of the parent and daughter atoms are denoted as $B_{p}$ (9.78855(25) eV) and $B_{d}$ (9.752390(15) eV)~\cite{NIST_ASD}.
%$B_{p}$ (9.78855(25) eV) and $B_{d}$ (9.752390(15) eV) are the electron binding energies of the parent and daughter atoms~\cite{NIST_ASD}.
%, which is neglected as it is on the order of few eVs and $R$ is close to 1. 
Since both the parent and daughter, as mass doublets, have the same $A/q$, the mass-dependent error becomes effectively negligible compared to the statistical uncertainty achieved in the measurement~\cite{Roux2013}. %Moreover, due to the fact that the mass difference of the parent and daughter is very small ($\Delta M/M$ < $10^{-4}$), the contribution of uncertainty to the $Q$-value from mass uncertainty of the reference (daughter), which amounts to 0.06 keV/c$^2$, can be disregarded.
Moreover, due to the fact that the mass difference of the parent and daughter is very small ($\Delta M/M$ < $10^{-4}$), the contribution from the reference (daughter) mass uncertainty of 0.06~keV/$c^2$ is suppressed by that factor and thus can be disregarded.
%77Ge -71212:87 0:05
%77Se -74599:50 0:06. -74599.49	0.062
%the mass-dependent error effectively becomes inferior compared to statistical uncertainty achieved in the measurements. 
%The determination of $Q$-value depends on the measured cyclotron  frequency $\nu_{c}$ via Eq.~\ref{eq:Qec}. 

%-----------------------------Table 1--------------------------------
%%%%%%%%%%%%%%%%%%%%%%% $^{75}$Ge (1/2$^{-}$) or $^{75}$Se  with ultra-low $Q$ values  initial 
\begin{table*}[!htb]
%\small
% \fontsize{7}{10}\selectfont 77As(3/2−)
   \caption{Transition from the ground state of the parent nucleus $^{77}$As  to the excited state of the daughter $^{77}$Se. The first  and second columns illustrate the experimental spin-parity of the initial ground state and its half-life. The third column gives the measured spin-parity of the excited final state. The fourth  column gives the decay type, which in this case is first-forbidden non-unique (1$^{\rm st}$ FNU).  The fifth column gives the gs-to-gs decay $Q$-value ($Q_{\beta^-}^0$)  from AME2020~\cite{Wang2021,Huang2021} and the sixth column from this work. The seventh and eighth columns give the gs-to-es decay $Q$ value ($Q_{\beta^-}^*$) from the literature~\cite{Wang2021} and this work, respectively. The ninth column lists the level of being positive/negative (in $\sigma$). The last  column gives the excitation energy E$^{*}$ from~\cite{NNDC}. All the energy values are in units of keV. 
%Spin-parity assignments enclosed in braces indicate that these are uncertain, which results in an uncertainty in the decay type, indicated by a  \{?\}. FNU denotes forbidden non-unique.
 %1st FNU represents 1st forbidden non-unique.
 }
 %in units of keV  with the experimental error   in units of keV.2nd FU represents 2nd forbidden unique
  \begin{ruledtabular}
   \begin{tabular*}{\textwidth}{@{}cccccccccc@{}}%{lcccccccc}%Initial state  & 
%\hline\hline
%\toprule
Initial state & Half-life&Final state &Decay type & \makecell[c]{$Q_{\beta^-}^0$ \\(AME2020)} &\makecell[c]{$Q_{\beta^-}^0$ \\ (This work)}& \makecell[c]{$Q_{\beta^-}^*$ \\(AME2020)} &\makecell[c]{$Q_{\beta^-}^*$ \\ (This work)}& \makecell[c]{$Q/\delta Q$ \\(This work)} & $E^{*}$ \\
\hline\noalign{\smallskip} 
%\midrule
       $^{77}$As (3/2$^{-}$)& 38.790(50) h&   $^{77}$Se (5/2$^{+}$)& 1st FNU &683.2(17) &684.463(70) &   3.1(17) &4.360(70)&62&680.1035(17)\\
   \end{tabular*}
   \label{table:low-Q}
   \end{ruledtabular}
\end{table*}
%-----------------------------Table 1 --------------------------------

\section{Results and discussion}

\begin{figure}[!htb]
   %\flushleft
   \includegraphics[width=0.99\columnwidth]{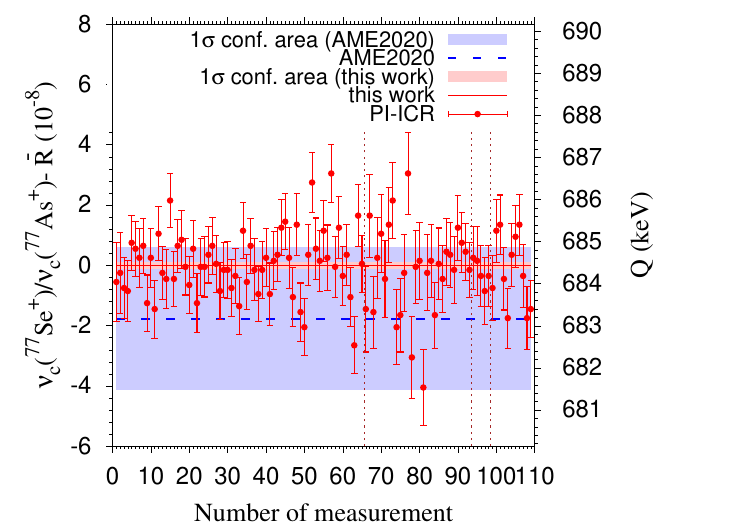}
   \caption{(Color online). The measured cyclotron frequency ratios  $R$ ( $\nu_c$($^{77}$Se$^{+}$)/$\nu_c$($^{77}$As$^{+}$)) (left axis) and $Q$ value (right axis) in this work compared to evaluated values from AME2020. The red dots with uncertainties are the measured PI-ICR single ratios in four time slots, which are separated with vertical brown dashed lines. The weighted average value in this work  $\overline{R}$ = 1.000 009 552 87(98) is represented by the solid red line and its 1$\sigma$ uncertainty band is shaded in red. The dashed blue line indicates the value adopted from AME2020  with its 1$\sigma$  uncertainty area shaded in blue.}
   \label{fig:ratio}
\end{figure}

%a grand total of 
The data collection involved initiating a $\nu_c$ measurement of $^{77}$Se$^{+}$ for four full scan rounds (one round consisting of 5$\times$5 points for both magnetron and cyclotron phases), followed by the measurement of $^{77}$As$^{+}$ for four full scan rounds. Subsequently, a center spot was recorded with $^{77}$Se$^{+}$ ions. In total, these steps lasted about seven minutes and were repeated.
% over a period of around 13 hours.
For each repetition, the positions of each spot were determined using the maximum likelihood method and the phase angles were calculated to deduce the cyclotron frequencies~\cite{nesterenko2021,Nesterenko2018,ge2021}.. 
The consecutive fitted cyclotron frequencies of $^{77}$Se$^{+}$ were linearly interpolated to the time of the measurement of $^{77}$As$^{+}$. This interpolated frequency was used to deduce the cyclotron resonance frequency ratio $R$. 
%In this manner, a total of about 70 frequency ratios were obtained.
%
The contribution of temporal fluctuations of the magnetic field to the final frequency ratio uncertainty was less than 10$^{-10}$ since the frequency measurements of the ion pair were tightly interleaved.
The incident ion rate was limited to a maximum of 5 detected ions/bunch with the median value being around 2 ions/bunch. Bunches containing more than 5 ions were excluded from the analysis to mitigate a potential cyclotron frequency shift due to ion-ion interactions~\cite{Kellerbauer2003,Roux2013}. Count-rate class analysis~\cite{Kellerbauer2003} was employed to verify that the frequency indeed did not shift.
By maintaining a small angle $\alpha_c$ $< 10$ degrees during the measurements, the frequency shifts due to ion image distortions remained well below the statistical uncertainty. 
The total measurement period spanned 13.1 hours, which we divided into four time slots: 7.2, 3.7, 0.6, and 1.6 hours as shown in Fig.~\ref{fig:ratio}. This division allowed us to maintain the desired small angle of $\alpha_c$. 
The weighted mean ratio $\overline{R}$ of the single ratios was computed alongside the inner and outer errors. The maximum of the inner and outer errors of the Birge ratios~\cite{Birge1932}, was adopted as the weights to calculate the final weighted mean cyclotron frequency ratio $\overline{R}$. 
The final frequency ratio $\overline{R}$ and the resultant $Q^0_{\beta^-}$-values are 1.000 009 552 87(98)  and 684.463(70)~keV, respectively. 
In Fig.~\ref{fig:ratio},  the results of the analysis are contrasted with the literature values.
%results of the analysis compared to the literature value are demonstrated. 

%-----------------------------Fig. 2 --------------------------------
%-----------------------------Fig. 2 --------------------------------

%-----------------------------Fig. 3 -------------------------------- 
\begin{figure}[!htb]
   %\flushleft
   \includegraphics[width=0.95\columnwidth]{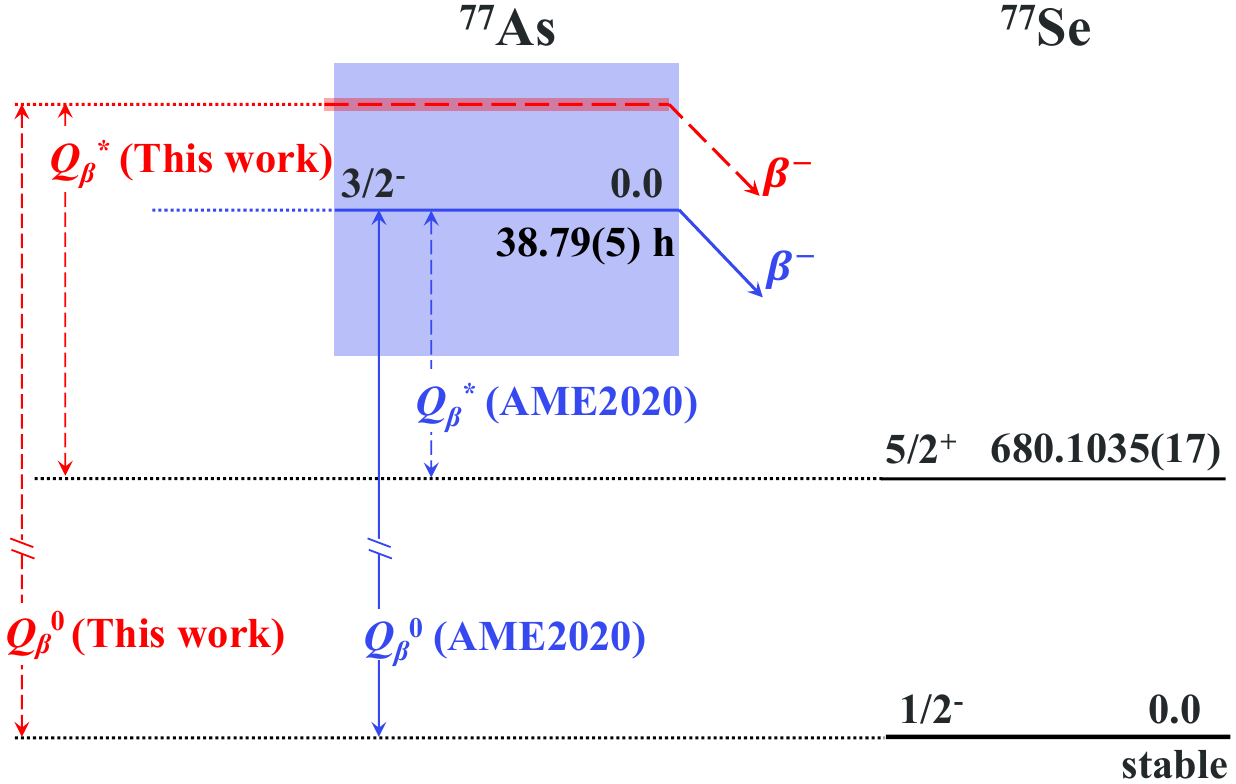}
   \caption{(Color online). The transition of $^{77}$As ground-state $\beta^-$ decay to the 680.1035(17)~keV $5/2^-$ state in $^{77}$Se. The horizontal blue line depicts the level with the $Q^0_{\beta^-}$ taken from AME2020~\cite{Huang2021,Wang2021} (shaded area shows the $1\sigma$ uncertainty) and the red dashed line the $Q^0_{\beta^-}$ from this work.
   The data for the  level scheme are adopted from~\cite{NNDC}.}
   \label{fig:level-scheme}
\end{figure}
%-----------------------------Fig. 3 --------------------------------

%-----------------------------Table 2 --------------------------------

 \begin{table*}[!htb]
\centering
\caption{Comparison of the experimental and \textit{jj44bpn} and \textit{jun45pn}-computed state energies $E_{\rm exc}$ (in units of MeV), electric quadrupole moments $Q_e$ (in units of barn), and magnetic dipole moments $\mu$ (in units of nuclear magneton $\mu_N$). The experimental evaluation data are from \cite{NNDC}. Effective charges adopted are $e_{\rm eff}^p$ = 1.5e and $e_{\rm eff}^n$ = 0.5e and the bare g-factors are $g_l(p)$ = 1, $g_l(n)$ = 0, $g_s(p)$ = $5.585$, and $g_s(n)$ = $-3.826$.}
\label{table:EM-observables}
\begin{ruledtabular}
\begin{tabular}{lccc|ccc|ccc}
\multicolumn{4}{c|}{Experimental Evaluation} & \multicolumn{3}{c|}{jj44bpn} & \multicolumn{3}{c}{jun45pn} \\
\cmidrule(lr){1-4} \cmidrule(lr){5-7} \cmidrule(lr){8-10}
Isotope ($J^\pi$) & $E_{\rm exc}$ (MeV) & $Q_e$ (barn) & $\mu$ ($\mu_N$) & $E_{\rm exc} $ (MeV) & $Q_e$ (barn) & $\mu$ ($\mu_N$) & $E_{\rm exc}$ (MeV)& $Q_e$ (barn) & $\mu$ ($\mu_N$) \\
\midrule
$^{77}$As $(3/2^{-})$ & 0.000 & - & +1.2940(13) & 0.000 & +0.2626 & +1.4775 & 0.000 & +0.2447 & +1.9403 \\
$^{77}$As $(5/2^{-})$ & 0.264 & - & +0.74(2) & 0.265 & -0.0899 & +0.9456 & 0.212 & -0.0488 & +0.4737 \\
$^{77}$Se $(1/2^{-})$ & 0.000 & - & +0.53356(5) & 0.000 & - & +0.5667 & 0.168 & - & +0.5959 \\
$^{77}$Se $(5/2^{-})$ & 0.250 & +0.76(5) & +1.12(3) & 0.486 & -0.1378 & +0.4319 & 0.337 & +0.376 & +1.3252 \\
$^{77}$Se $(5/2^{-})$ & 0.439 & - & +1.0(3) & 0.665 & +0.2128 & +1.2053 & 0.677 & -0.1671 & +0.7419 \\
$^{77}$Se $(5/2^{+})$ & 0.680 & - & - & 0.561 & -0.3902 & -0.9399 & 0.635 & -0.4334 & -1.0071
\end{tabular}
\end{ruledtabular}
\end{table*}
%-----------------------------Table 2 --------------------------------

%The final $Q^0_{\beta^-}$-value and the mass-excess of $^{77}$As obtained from the mean cyclotron frequency ratio is given in  Table~\ref{table:low-Q}. 
The $Q^0_{\beta^-}$ value of $^{77}$As from this work is a factor of 24 more precise and 1.3(17)~keV larger than the value in AME2020~\cite{Huang2021}. % The value in AME2020   ${\beta^{+}}$-decay measurement of $^{77}$As(${\beta^{+}}$)$^{77}$Se. % This was said already earlier (I think) 
The mass-excess value of $^{77}$As was deduced to be -73915.026(94)~keV/c$^2$ and the precision was improved by a factor of 18.  The uncertainty of the mass excess for $^{77}$As has an additional 0.06~keV/c$^2$ uncertainty in the  mass of the daughter $^{77}$Se as reference, which was evaluated based on (n,$\gamma$) reaction experiments ~\cite{ENGLER1981,TOKUNAGA1985,osti88} in AME2020~\cite{Huang2021,Wang2021}. 
%It has an additional 0.08~keV/c$^2$ uncertainty contribution from the uncertainty of the reference daughter $^{77}$Se ion mass.
Combining the new $Q^0_{\beta^-}$ value together with the nuclear energy level data gives the final $Q$-values for decays to the potential low $Q$-value states, as tabulated in Table \ref{table:low-Q}  and  illustrated in Fig.~\ref{fig:level-scheme}. 
The gs-to-es $Q^*_{\beta^-}$ value of 4.360(70) keV for the decay channel $^{77}$As (3/2$^{-}$) $\rightarrow$ $^{77}$Se (5/2$^{+}$, 680.1035(17) keV),  is comparable to the  presently running  direct (anti)neutrino mass experiments using 
%$\beta^{-}$  decaying  $^{187}$Re with a $Q$ value of 2.4709(13) keV~\cite{Filianin21} and 
electron-capture decaying $^{163}$Ho with a $Q$ value of  2.8632(6) keV~\cite{schweiger2024}.
%2.833(30)$_\textrm{stat}$(15)$_\textrm{sys}$ keV~\cite{Eliseev2015}.
% 2.8632(6) keV~\cite{schweiger2024}
%%2.492(30)$_\textrm{stat}$(15)$_\textrm{sys}$~KeV
%change
%$\beta^{-}$  decaying  $^{187}$Re with a $Q$ value of 2.4709(13) keV~\cite{Filianin21} and electron-capture decaying $^{163}$Ho with a $Q$ value of 2.833(30)$_\textrm{stat}$(15)$_\textrm{sys}$ keV~\cite{Eliseev2015}.
%to electron-capture decaying $^{163}$Ho with a $Q$ value of  2.8632(6) keV~\cite{schweiger2024}.

The transition of interest is first-forbidden non-unique (1$^{\rm st}$ FNU) and thus depends on nuclear structure through nuclear matrix elements (NME). To compute these NME and to
estimate the half-life of this transition, nuclear shell-model (NSM) calculations utilizing the software $\textit{KSHELL}$~\cite{Shimizu2019}, with the well established effective interactions \textit{jj44bpn}~\cite{Honma2009} and \textit{jun45pn}~\cite{Mukhopadhyay2017}, were performed. Their model spaces consist of adopting $^{56}$Ni as a closed core with the orbitals 0f$_{5/2}$, 1p$_{3/2}$, 1p$_{1/2}$, and 0g$_{9/2}$ for both protons and neutrons. We assess the reliability of the nuclear Hamiltonians employed by analyzing the magnetic dipole and electric quadrupole moments, alongside the excitation energies, predicted by the models in Table \ref{table:EM-observables}. The computed electromagnetic properties and the excitation energies are generally in fair agreement with the evaluation data. 

\begin{figure}[!htb]
   %\flushleft
   \includegraphics[width=0.995\columnwidth]{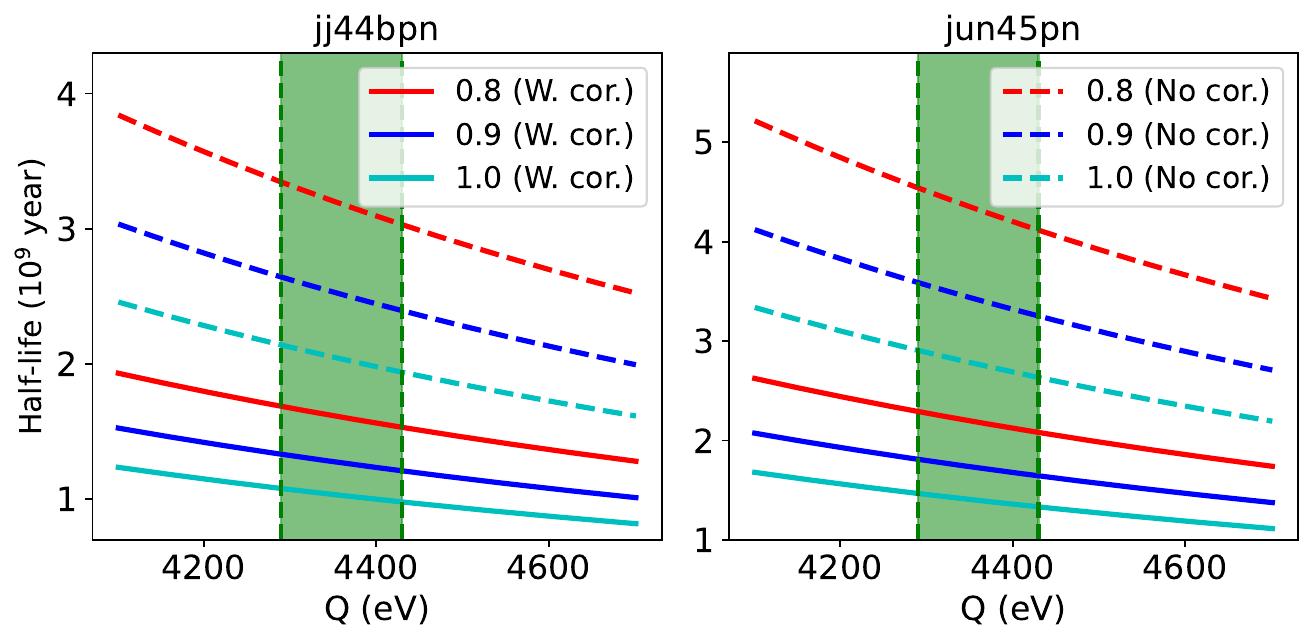}
   \caption{(Color online). Computed partial half-lives as a function of available energy ($Q$ value) for the transition $^{77}$As $(3/2^{-})$ $\rightarrow$ $^{77}$Se $(5/2^{+})$ for different choices of the axial-vector coupling $g_{\rm A}$ (0.8, 0.9 and 1.0) using the effective Hamiltonians \textit{jun45pn} (left panel)  and \textit{jun45pn}  (right panel). The legend includes half-life values with (W.) and without (No) the atomic exchange correction (cor.). Additionally, green vertical dashed lines shaded in between with green color denote the highest and lowest available energies derived in this work for the transition, which are 4.430 keV and 4.290 keV, respectively.
   %Additionally, black vertical dashed lines denote the highest (4.430 keV) and lowest (4.290 keV) available energies from this work. The energy range derived in this work is shaded in green.
   % for the transition 
   %The parentheses in the legend  give the half-life values for computations with (W.) or without (No) the atomic exchange correction (cor.). The  vetical dash lines in black indicate the highest (lowest) available energy, 4.430 (4.290) keV, for the transition. 
   %lower (upper) half-life columns correspond to the 
   }
   \label{fig:half-lives}
\end{figure}

The dependence of the partial half-life of the 3/2$^{-} \rightarrow 5/2^+$ transition on the $Q$ value is shown in Fig.~\ref{fig:half-lives}. The computations of the $\beta^-$ transition rate account for screening, radiative, and atomic exchange corrections. The atomic exchange correction was originally derived for allowed $\beta$ decays by Nitescu \textit{et al.} \cite{Nitescu2023} and is the most important contribution due to the low $Q$ value of the discussed transition, as can be seen in Fig. ~\ref{fig:half-lives}. The computed half-life is of the order of 10$^{9}$ years for this 1$^{\rm st}$ FNU transition, which rules out this candidate to be a potential source for rare-event experiments searching for the electron antineutrino mass.
Due to the short half-life (less than 2 days) of the isotope $^{77}$As, it poses additional  challenges for inclusion in a long-term neutrino mass determination experiment.
%With a half-life of less than 2 days of the isotope, $^{77}$As, makes it difficult to be realized in a long-term neutrino mass determination experiment. 
%is shorter than that of $\beta^{-}$  decaying  $^{187}$Re, being used for antineutrino-mass determination~\cite{Nesterenko2014}, with a half-life of 4.23(13)$\times$10$^{10}$ years~\cite{LINDNER19891597}.

%-----------------------------Fig. 4 -------------------------------- 
%-----------------------------Fig. 4 --------------------------------

\section{Conclusion and Outlook}
A direct high-precision gs-to-gs $^{77}$As($\beta^-$)$^{77}$Se decay $Q$-value measurement was performed using the PI-ICR technique at the JYFLTRAP Penning trap mass spectrometer. A $Q$ value of 684.463(70)~keV was obtained from the cyclotron frequency ratio measurements to a relative precision better than 1$\times 10^{-9}$ of the ions of the decay pair.
%A discrepancy of more than three standard deviations was found to the previously adopted value in the AME2020.
The refined $Q$-value is in good agreement with the evaluated value in AME2020 but its uncertainty was improved by a factor of 24. The candidate FNU transition $^{77}$As (3/2$^{-}$)$\rightarrow ^{77}$Se (5/2$^{+}$, 680.1035(17) keV) is confirmed to be energetically allowed at a level of more than 60$\sigma$. Furthermore, the newly derived gs-to-es $Q$ value allowed us to quantify the order of magnitude of the partial half-life of this transition by using nuclear wave functions obtained with the use of two different established Hamiltonians in nuclear shell-model calculations. 
The obtained half-life of approximately  $10^9$ years for this 1$^{\rm st}$ FNU transition, excludes $^{77}$As as a viable candidate for future long-term neutrino mass determination experiments.

%The obtained half-life was found to be around $10^9$ years for this 1$^{\rm st}$ FNU transition, which eliminates $^{77}$As as a potential candidate for future long-term neutrino mass determination experiments.
%The obtained half-life turned out to be of the order of $10^9$ years, leading to an exclusion of $^{77}$As as candidate for future long-term neutrino mass determination experiment.
%beta-decay half-life of 187Re to be (4.23 ± 0.13) · 1010 y
%The obtained half-life of approximately 10^9 years for the first FNU transition excludes 77As as a viable candidate for future long-term neutrino mass determination experiments.
%, thus verifying it as a possible candidate for future antineutrino-mass determination experiments
%\section{Acknowledgements}

%\section{Acknowledgements}
\acknowledgments 
We acknowledge the staff of the Accelerator Laboratory of University of Jyv\"askyl\"a (JYFL-ACCLAB) for providing stable online beam. We thank the support by the Academy of Finland under the Finnish Centre of Excellence Programme 2012-2017 (Nuclear and Accelerator Based Physics Research at JYFL) and projects No. 306980, No. 312544, No. 275389, No. 284516, No. 295207, No. 315179, No. 327629, No. 354589, No. 345869, and No. 354968. The support by the EU Horizon 2020 research and innovation program under grant No. 771036 (ERC CoG MAIDEN) is acknowledged.  This project has received funding from the European Union’s Horizon 2020 research and innovation programme under grant agreement No. 861198–LISA–H2020-MSCA-ITN-2019. This project has received funding from the European Union’s Horizon Europe Research and Innovation Programme under Grant Agreement No. 101057511 (EURO-LABS).
% We acknowledge the staff of the accelerator laboratory of University of Jyv\"askyl\"a (JYFL-ACCLAB) for providing stable online beam and J.~Jaatinen and R.~Sepp\"al\"a for preparing the production target. We thank the support by the Academy of Finland under the Finnish Centre of Excellence Programme 2012-2017 (Nuclear and Accelerator Based Physics Research at JYFL) and projects No. 306980, 312544, 275389, 284516, and 295207. The support by the EU Horizon 2020 research and innovation program under grant No. 771036 (ERC CoG MAIDEN) is acknowledged.
 % Academy of Finland, Grant Nos. 314733 and 320062
 
%\blindtext~\cite{article-minimal}

\bibliographystyle{apsrev4-1}

%\bibliography{My_Collection_merged_with_Tommi}
%\bibliography{tommi_bib_2020-10-03}
\bibliography{my-final-bib-from-jabref}
%\bibliography{my-final-bib-from-jabref,ref}
%\bibliography{My Collection}
%\bibliography{final-mendeley-merged} 
\end{document}